\documentclass[
    aps,
    prd,
    reprint,
    superscriptaddress
]{revtex4-2}
\usepackage{mathtools}
\usepackage{amsmath}
\usepackage{amsbsy}
\usepackage{graphicx}
\usepackage[colorlinks=true,linkcolor=blue,citecolor=blue,urlcolor=blue]{hyperref}

\def\eos#1{equation of state#1 (EOS#1)\gdef\eos{EOS}}
\def\QNM#1{quasi-normal mode#1 (QNM#1)\gdef\QNM{QNM}}
\def\ns#1{neutron star#1 (NS#1)\gdef\ns{NS}}
\def\gw#1{gravitational wave#1 (GW#1)\gdef\gw{GW}}
\def\bh#1{black hole#1 (BH#1)\gdef\bh{BH}}
\def\bbh#1{binary black hole#1  (BBH#1)\gdef\bbh{BBH}}
\def\bns#1{binary neutron star#1 (BNS#1)\gdef\bns{BHS}}
\def\bhns#1{black hole - neutron star#1 (BHNS#1)\gdef\bhns{BHNS}}
\def\nsbh#1{neutron star - black hole#1 (NSBH#1)\gdef\nsbh{NSBH}}
\def\tov#1{Tolman-Oppenheimer-Volkoff#1 (TOV#1)\gdef\tov{TOV}}
\def\nr#1{numerical relativity#1 (NR#1)\gdef\nr{NR}}
\def\ligo#1{Laser Interferometer Gravitational-wave Observatory#1 (LIGO#1)\gdef\ligo{LIGO}}

\newcommand{\whisky}{\texttt{Whisky} }
\newcommand{\etk}{\texttt{Einstein Toolkit} }

\begin{document}

\title{Relativistic Simulations of Self-Gravitating Radiation Fluids}
\author{Bing-Jyun Tsao}
\affiliation{Center for Gravitational Physics, Weinberg Institute,
Department of Physics, The University of Texas at Austin,
Austin, TX 78712, USA}
\affiliation{Cornell Center for Astrophysics and Planetary Science,
Cornell University, Ithaca, New York 14853, USA}

\author{Pablo Laguna}
\affiliation{Center for Gravitational Physics, Weinberg Institute,
Department of Physics, The University of Texas at Austin,
Austin, TX 78712, USA}

\author{Octavio Guerrero}
\affiliation{Center for Gravitational Physics, Weinberg Institute,
Department of Physics, The University of Texas at Austin,
Austin, TX 78712, USA}

\date{\today}

\begin{abstract}
    We present an approach for numerical simulations of pure radiation fluids within the general relativistic hydrodynamics framework. The approach models radiation as a self-gravitating perfect fluid with vanishing rest mass, enabling simulations of radiation fluids in strong-gravity spacetimes. Building on the existing relativistic hydrodynamics code \texttt{Whisky}, within the \texttt{Einstein Toolkit} framework, we introduce modifications to incorporate radiation fluids, including updates to the source terms, recovery scheme, and Riemann solvers. Modifications are validated using standard shock tube tests. We present results for radiation accretion onto a single black hole, as well as a photon star solution for radiation in unstable hydrostatic equilibrium, demonstrating both collapse and dispersion under perturbations.
\end{abstract}
\maketitle

\section{Introduction}
    Relativistic hydrodynamical methods have been widely explored for applications of simulating jets and compact binaries. Several relativistic hydrodynamical codes are supported within the \etk{} framework, including the \whisky code~\cite{Baiotti:2003btu,Hawke:2005zw,Baiotti:2004wn}, \texttt{GRHydro}~\cite{GRHydro_2013}, and \texttt{IllinoisGRMHD}~\cite{Etienne:2015cea}. These codes solve the fluid evolution equations in flux-conservative forms while implementing reconstruction and shock-capturing schemes for finite volume method. In this work, we present a simple modification to the existing relativistic hydrodynamics code to evolve self-gravitating radiation fluid with vanishing baryon mass to simulate spacetime containing radiation but no baryon matter.

    Radiation transport coupled to relativistic hydrodynamics has been widely studied in astrophysical settings using a variety of numerical methods to approximate the solutions to the Boltzmann equation~\cite{Lindquist_1966,Zanotti:2011mb,Sekiguchi:2011zd,Ruffert:1995fs}. Common approaches include leakage schemes~\cite{Sekiguchi:2010fh}, which approximate radiation emission and cooling without explicitly evolving the radiation field; truncated-moment schemes~\cite{Thorne:1981nvt,Shibata:2011kx,Radice:2021jtw}, which evolve moments of the radiation distribution function and provide a closure for the unresolved higher moments; Monte Carlo methods~\cite{Foucart_2020,Foucart:2021mcb,Ryan:2015mda,Miller:2019gig}, which represent the radiation field using discrete packets that propagate along geodesics and interact with matter probabilistically; and hybrid methods that evolve moments with Monte-Carlo closures~\cite{Foucart:2017mbt}. These methods have been applied to both photon and neutrino transport and allow the radiation to interact with fluid matter through emission, absorption, and scattering. In this work, we instead consider a simplified description in which the radiation itself is treated as a perfect fluid with the $\Gamma$-law equation of state $P=(\Gamma-1)e$ in the ultra-relativistic limit $\Gamma=4/3$ and vanishing rest mass density.
    
    The perfect-fluid treatment of radiation provides a framework for studying the dynamical evolution of self-gravitating photons in strong-field spacetimes. The entropy of self-gravitating radiation has been studied in~\cite{Sorkin_1981}, where spherically symmetric, static equilibrium configurations within a confined radius are shown to be extrema of the total entropy. These ``photon star'' solutions~\cite{Schmidt_2000} can be divided into two families by integrating the \tov{} equations inward from a fixed radius~\cite{Anastopoulos:2011av,Anastopoulos:2020mrt}: regular solutions with vanishing central mass and singular solutions with negative central mass. The stability and thermal properties of singular \tov{} solutions were studied in~\cite{Laguna:2026hfa,Zurek:1984,tHooft:1997iys}. Although regular \tov{} solutions exist, small perturbations can lead to collapse as $\Gamma=4/3$ fluids are critical solutions to the \tov{} equations. Critical collapse has been studied for massless scalar fields~\cite{Choptuik_1993}, gravitational-wave packets~\cite{Abrahams_1993}, and self-gravitating radiation~\cite{Evans:1994pj}. To demonstrate the application of the radiation fluid code, we evolve a regular solution and show its dynamical instability in Sec.~\ref{sec:Tests}.

    The paper is organized as follows: the initial data and evolution methods are described in Sec.~\ref{sec:GRHD}. Numerical implementations and modifications for radiation fluids are described in Sec.~\ref{sec:NR}. Results from the numerical tests are discussed in Sec.~\ref{sec:Tests}, with conclusions in Sec.~\ref{sec:Conclusions}. The methods and results are reported in units of $G = c = 1$, with mass scale in the code $M=1$. The space-time signature is $(-,+,+,+)$, and tensor indices are denoted with Greek letters $(\mu,\nu...)$. Spatial tensor indices are denoted with Latin letters from the middle of the alphabet $(i,j,k...)$.

\section{General relativistic hydrodynamics for radiation}
    \label{sec:GRHD}
    For full general relativistic spacetime evolution, we follow the 3+1 decomposition of the metric in the Arnowitt–Deser–Misner (ADM) formalism~\cite{Arnowitt:1962hi}, where spacetime is foliated into 3D spatial hypersurfaces. In this decomposition, the line element is written as
    \begin{align*}
        ds^2 &= g_{\mu\nu} dx^\mu dx^\nu \\
             &= -\alpha^2 dt^2 + \gamma_{ij}\left(dx^i + \beta^i dt\right)\left(dx^j + \beta^j dt\right),
    \end{align*}
    with $g_{\mu\nu}$ being the spacetime metric and $\alpha,\beta^i,\gamma_{ij}$ being the lapse, shift vector, and spatial 3-metric induced on the hypersurface. We consider fluid with 4-velocity $u^\mu$. The Lorentz factor is defined as
    \begin{equation*}
        W = \alpha u^t = \left(1-\gamma_{ij} v^i v^j\right)^{-1/2},
    \end{equation*}
    where the fluid 3-velocity is given by
    \begin{equation*}
        v^i = \frac{u^i}{\alpha u^t} + \frac{\beta^i}{\alpha}.
    \end{equation*}

For baryon fluids, the hydrodynamics variables are related through
\begin{equation*}
    e = \rho_0+\rho_0 \epsilon, \quad h= 1 + \epsilon + \frac{P}{\rho_0}
\end{equation*}
where $\rho_0, \epsilon, e, h, P$ are the rest mass density, specific internal energy, total energy density, specific enthalpy, and pressure, respectively. Since radiation has no rest mass $\rho_0 \rightarrow{} 0$, we evolve the total energy density $e$ instead of $\rho_0$ and $\epsilon$. Specifically, we perform the following replacements to the hydrodynamics variables:
\begin{equation*}
    \rho_0 \epsilon \rightarrow{} e, \quad \rho_0 h \rightarrow{} e+P
\end{equation*}
The $\Gamma$-law equation of state $P(e) = (\Gamma - 1) e$ where $\Gamma = 4/3$ is used for ultra-relativistic fluids. For this equation of state, the sound speed of the radiation fluid is constant
\begin{equation*}
    c_s^2 = \left.\frac{\partial P}{\partial e}\right|_s = \Gamma - 1 = \frac{1}{3}.
\end{equation*}

Given the above substitutions for the hydro variables, we can model the radiation as a perfect fluid with stress-energy tensor
\begin{equation}
    T^{\mu\nu} = (e + P) u^\mu u^\nu + P g^{\mu\nu}.
\end{equation}
The projections of the stress-energy tensor are defined as
\begin{align}
    \label{eq:T_proj}
    \rho_H &= T_{\mu\nu} n^\mu n^\nu
       = (e+P)W^2 - P,\\
    J_i &= -T_{\mu\nu} n^\mu \gamma^\nu{}_{i}
       = (e+P)W^2 v_i ,\\
    S_{ij} &=
    T_{\mu\nu}\gamma^\mu{}_{i}\gamma^\nu{}_{j}
    = (e+P)W^2 v_i v_j + P\gamma_{ij}.
\end{align}

We evolve the relativistic hydrodynamic equations in a 3+1 Valencia formulation, which recasts the conservation of rest mass density and stress-energy
\begin{equation}
    \nabla_\mu(\rho_0 u^\mu) = 0, \qquad
    \nabla_\mu T^{\mu\nu} = 0
\end{equation}
into a flux-conservative form suitable for high-resolution shock-capturing schemes, 
\begin{equation}
    \partial_t \mathbf{U} + \partial_i \mathbf{F}^i = \mathbf{S}.
\end{equation}

For baryon fluids, there are 5 conservative variables $\{D, S_j, \tau\}$ evolved in the Valencia formulation:
\begin{align*}
    D &\equiv \sqrt{\gamma} \rho_0 W \\
    S_j  &\equiv \sqrt{\gamma} (e+P) W^2 v_j \\
    \tau &\equiv \sqrt{\gamma} ((e+P) W^2 - P) , 
\end{align*}
where $\gamma$ is the determinant of the spatial metric $\gamma_{ij}$. For radiation fluids, the conservation of rest mass energy is trivial and the first term $D = \sqrt{\gamma} \rho_0 W= 0$ vanishes, which leaves only $4$ conservative variables $\{S_j, \tau\}$. 

\begin{equation}
    \mathbf{U} =
    \begin{pmatrix} 0 \\ S_j \\ \tau  \\
    \end{pmatrix}
    = \sqrt{\gamma}
        \begin{pmatrix} 0 \\
            (e+P) W^2 v_j \\
            ((e+P) W^2 - P) \\ 
        \end{pmatrix},
\end{equation}
\begin{equation}
    \mathbf{F}^i = \alpha\sqrt{\gamma}
    \begin{pmatrix}
    0 \\
    (e+P) W^2 \tilde{v}^i v_j + P \delta^i_j\\
    \left((e+P) W^2 - P\right) \tilde{v}^i + P v^i \\ 
    \end{pmatrix},
\end{equation}
\begin{equation}
    \mathbf{S} = \alpha \sqrt{\gamma}
    \begin{pmatrix}
        0 \\
        T^{\mu\nu}\left(\partial_\mu g_{\nu j} - \Gamma^\sigma_{\mu\nu} g_{\sigma j}\right) \\
        T^{\mu 0} \partial_\mu \alpha  - \alpha T^{\mu\nu} \Gamma^0_{\mu\nu} \\
    \end{pmatrix}.
\end{equation}
where $\tilde{v}^i \equiv v^i - \beta^i/\alpha$ and $\Gamma^\sigma_{\mu\nu}$ are the Christoffel symbols of the 4-metric $g_{\mu\nu}$.
\section{Numerical Implementation}
    \label{sec:NR}
    The coupled spacetime and hydrodynamics equations are solved with the \texttt{MAYA} code~\cite{Evans_2015,Clark_2016,Jani_2016}, our local version of the \texttt{Einstein Toolkit} code~\cite{EinsteinToolkit:2023_11}. Einstein equations are solved with the BSSN-Chi formulation~\cite{Shapiro1999,Shibata1995,Beyer:2004sv}. To evolve the spacetime, we use the moving puncture gauge~\cite{Campanelli2005,Baker2006}. The initial data for the spacetime variables are constructed by solving the Hamiltonian constraint using a modified \texttt{TwoPunctures}~\cite{Ansorg_2004} code to include the source term from the energy density of the fluid.

    The general relativistic hydrodynamics evolution for radiation is modified the \texttt{Whisky} public code~\cite{Baiotti:2004wn,Hawke:2005zw, Baiotti:2003btu}. The code uses a high-resolution shock-capturing finite-volume method to evolve the cell-averaged fluid variables. For simplicity, the discretization of the update in the $x$-direction is
    \begin{equation}
        \frac{d \mathbf{U}_i}{dt}
        =
        -\frac{\hat{\mathbf{F}}^x_{i+1/2}-\hat{\mathbf{F}}^x_{i-1/2}}{\Delta x}
        +
        \mathbf{S}_i .
    \end{equation}
    where the numerical flux $\hat{\mathbf{F}}^x_{i+1/2}$ depends on the left and right reconstruction of the conservative variables $(\mathbf{U}^L_{i+1/2},\mathbf{U}^R_{i+1/2})$. The reconstruction on cell surfaces are performed with the Piecewise Parabolic Method (PPM)~\cite{PPM_1984}, and the Riemann solver to capture shocks is computed with the Harten-Lax-van Leer-Einfeldt (HLLE)~\cite{Harten:1983,Einfedldt_1988} and the Marquina solvers~\cite{Donat_1998,ibanez_1999}. The details for the modifications to the \whisky code for radiation fluid are discussed in this section.

    \subsection{Recovery}
        Similar to the primitive variable recovery procedure described in~\cite{GRHydro_2013}, we work with the undensitized conservative variables obtained by dividing out the metric determinant factor $\hat{\tau} = \tau/\sqrt{\gamma}$ and $\hat{S_j} = S_j/\sqrt{\gamma}$. The recovery of $\{e,W\}$ from $\{\hat{S_j},\hat{\tau}\}$ is given by a simple quadratic formula:
        \begin{align}
            W^2 &= \frac{-B + \sqrt{B^2-4AC}}{2A} \\
            e &= \frac{\hat{\tau}}{(\Gamma W^2 - (\Gamma -1))}
        \end{align}
        where
        \begin{align*}
            A &= \Gamma^2 (\hat{\tau}^2 - \hat{S}^2) \\
            B &= 2\Gamma(\Gamma-1) \hat{S}^2 - \Gamma^2 \hat{\tau}^2 \\
            C &= -\hat{S}^2 (\Gamma - 1)^2
        \end{align*}
        
        This procedure becomes numerically unstable at very high Lorentz factors as $\hat{S}^2 \rightarrow \hat{\tau}^2$, which happens near the punctures and also in low-density outflow. We follow the Faber's puncture trick proposed in~\cite{Faber_2007} and modified it for radiation fluid by imposing the upper limit for $\hat{S}^2$ by $0.98 \tilde{\tau}^2$ to avoid $A<0$. 

    \subsection{Riemann Solver}
        Although radiation has a constant sound speed $c_s^2 = 1/3$, the evolution equations remain a nonlinear hyperbolic system that generically develops shocks, requiring a Riemann solver to resolve the characteristic structure. In this study, we modify the HLLE and the Marquina solvers for radiation fluid to handle shocks.\\
        
        These formalisms perform a characteristic decomposition of the flux Jacobian for each direction $i=x,y,z$
        \begin{equation}
            \mathbf{A}^i = \frac{\partial \mathbf{F}^i}{\partial \mathbf{U}},
        \end{equation}
        to obtain the eigenvalues $\lambda_{(p)}$ and corresponding right and left eigenvectors $r_{(p)}$ and $l_{(p)}$, defined by
        \begin{equation}
            \mathbf{A}^i r_{(p)} = \lambda_{(p)} r_{(p)}, \qquad
            l_{(p)} \mathbf{A}^i = \lambda_{(p)} l_{(p)}.
        \end{equation}
        
        Among the four eigenvalues, $\lambda_0$ has a double degeneracy (rather than a triple in baryon fluids), while the other two $\lambda_\pm$ retain the same expression as in~\cite{ibanez_1999} (as an example, in the $x$-direction):
            \begin{align*}
            \lambda_0 &= \alpha v^x - \beta^x \text{ (double)}\\
            \lambda_\pm
            &=
            \alpha \frac{
            (1 - c_s^2)\, v^x
            \pm
            \mathcal{C}}{1 - c_s^2 v^2}
            -
            \beta^x
        \end{align*}
            where $\mathcal{C}=\sqrt{c_s^2 (1 - v^2)\bigl(\gamma^{xx} (1 - v^2 c_s^2)-(1 - c_s^2)v^x v^x\bigr)}$.

        Characteristic decomposition is performed for both the left and right sides of the interfaces given the primitive variables recovered from $\mathbf{U}^L, \mathbf{U}^R$. The Riemann solvers then construct the numerical flux by adding dissipation to the averaged physical fluxes of the left and right sides $\mathbf{F}^L = \mathbf{F}(\mathbf{U}^L), \mathbf{F}^R=\mathbf{F}(\mathbf{U}^R)$.
        
            \subsubsection*{HLLE flux}
                The HLLE Riemann solver evaluates the flux with a two-wave approximation given by the minimum and maximum characteristic speeds $\xi_-,\xi_+$:
                \begin{equation*}
                    \xi_- = \min_p \left( \lambda^L_{(p)}, \lambda^R_{(p)} \right), \qquad \xi_+ = \max_p \left( \lambda^L_{(p)}, \lambda^R_{(p)} \right),
                \end{equation*}
                The HLLE flux along the interface takes the form
                \begin{equation}
                    \hat{\mathbf{F}}(U) =
                    \frac{
                    \hat{\xi}_+ \mathbf{F}(\mathbf{U}^L)
                    - \hat{\xi}_- \mathbf{F}(\mathbf{U}^R)
                    + \hat{\xi}_+ \hat{\xi}_- (\mathbf{U}^R - \mathbf{U}^L)
                    }{
                    \hat{\xi}_+ - \hat{\xi}_-
                    },
                \end{equation}
                where $\hat{\xi}_- = \min(0,\xi_-), \ \hat{\xi}_+ = \max(0,\xi_+).$\\
                
            Since the HLLE solver does not use the eigenvectors, and the eigenvalues $\lambda_{(p)}$ remain the same expression as the baryon fluid, no modifications are needed for radiation fluid.
            
            \subsubsection*{Marquina flux}
                The Marquina Riemann solver routine evaluates the left and right eigenvectors in addition to the eigenvalues~\cite{Donat_1998,ibanez_1999}. The eigenvectors for baryon fluids depend on rest mass density and enthalpy, so it is expected that modified expressions will be needed for radiation fluids. The left and right eigenvectors are presented in Appendix~\ref{sec:appendix1_riemann}. The Marquina fluxes along the interfaces are given by
                \begin{equation}
                    \hat{\mathbf{F}}(\mathbf{U}) = \frac{1}{2}(\mathbf{F}^R + \mathbf{F}^L + \mathbf{Q}^R - \mathbf{Q}^L),
                \end{equation}
                where $Q^L, Q^R$ are the numerical viscosity terms,
                \begin{equation}
                    \mathbf{Q}_j^{(L,R)}
                    =
                    \sum_{(p),i=1}^{4}
                    |\lambda_{(p)}|_{\max(L,R)}
                    \, r_{(p)j}^{(L,R)} \, l_{(p)i}^{(L,R)} \, \mathbf{U}_i^{(L,R)}.
                \end{equation}

\section{Numerical tests}
    \label{sec:Tests}
    The numerical tests performed for the radiation fluids utilized the analysis tools in the \texttt{Einstein Toolkit} framework. The \bh{} is tracked with the \texttt{AHFinderDirect} code~\cite{Thornburg:2003sf}. The total energy of the radiation fluid is tracked via the modified \texttt{VolumeIntegrals} thorn~\cite{EinsteinToolkit:2023_11}. The mass, spins, and multipole moments of the \bh{} are computed using the \texttt{QuasiLocalMeasures} thorn~\cite{Dreyer:2003} based on the dynamical horizons framework~\cite{Ashetekar_2004}. The 2D visualization is made with the \texttt{kuibit}~\cite{kuibit} tool. 
    
    For the numerical tests, we employ a $4^{\mathrm{th}}$-order Runge--Kutta scheme for time integration with a Courant--Friedrichs--Lewy (CFL) number of $0.5$ at the finest resolution level. We employ the radiative boundary condition for both fluid and spacetime variables. Perturbations in the energy density is introduced using a Gaussian profile of the form
    \begin{equation*}
        \delta e(r) = \delta e_0 \exp\left(-\frac{(r - r_0)^2}{2\sigma^2}\right),
    \end{equation*}

\subsection{Standard shock test}
    \begin{figure}[!htbp]
        \centering
        \includegraphics[width=0.48\textwidth]{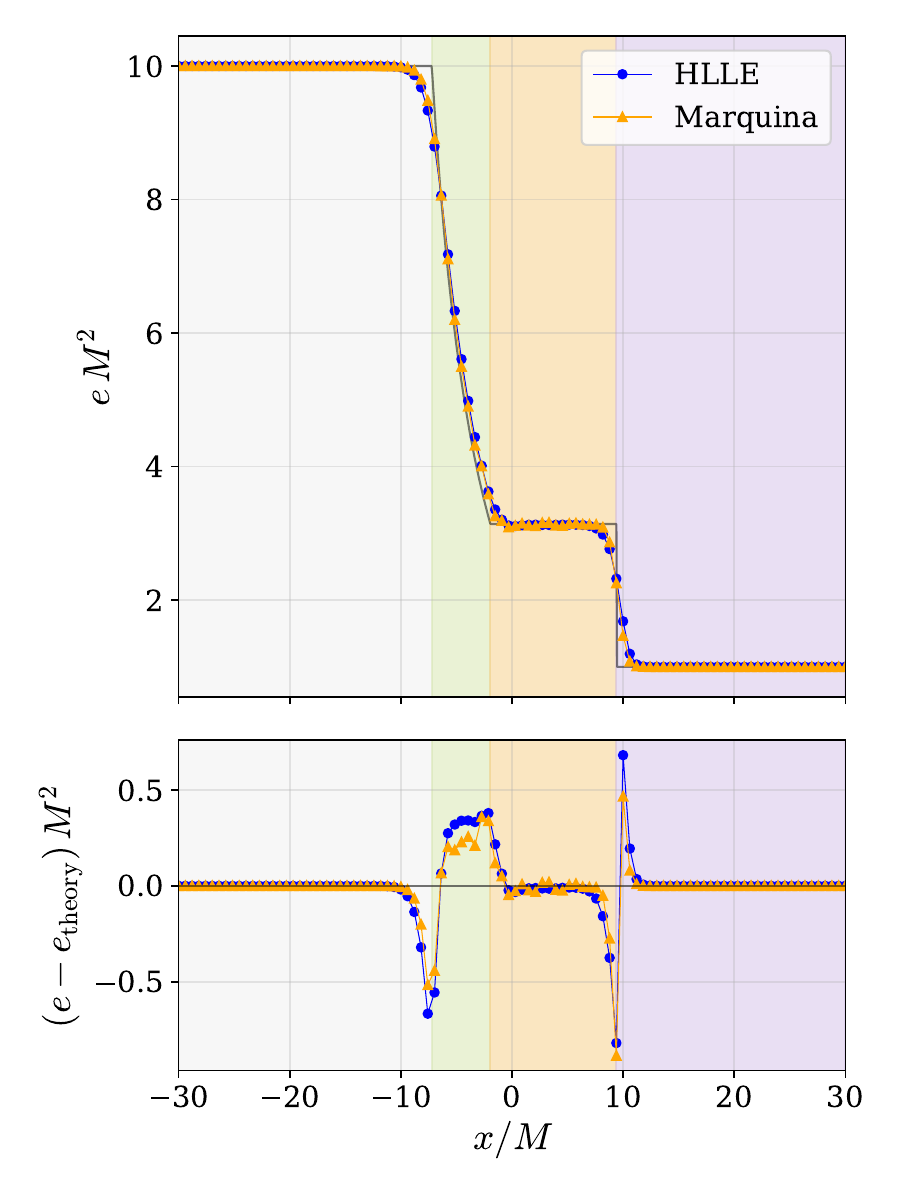}
        \caption{The shock tube test for the two Riemann solvers. The top plot shows the energy density from HLLE (blue circles) and Marquina (orange triangles) solvers compared to the theoretical solution (black curve), and the bottom plot shows the absolute errors of the two methods. Rarefaction wave from the high density region is shown in green, while the shock between the orange and blue region travels into the low density region.}
        \label{fig:shocktube}
    \end{figure}
    As a first test, we consider the standard shock tube test with the energy density (in units $M^{-2}$) discontinuity in the $x$-direction
    \begin{align*}
        x < 0, \qquad e_L &= 10, \quad     v_L = 0 \\
        x \geq 0, \qquad e_R &= 1,\quad \ \   v_R = 0
    \end{align*}

    The discontinuity is initially at $x=0$. The grid uses $140$ points in the $x$-direction, and $10$ points in $y,z$-directions, with resolution $\Delta=0.5M$.\\
    
    The results at iteration $100$ for the shock tube test are shown in Fig.~\ref{fig:shocktube} for both the HLLE (in blue) and Marquina (in orange) Riemann solvers. The top plot shows the rarefaction wave (green background) retracting from the high density region on the left (gray background), while the shock wave separating the orange and blue regions travels toward the low density region, showing good agreement with analytical results (in black). 

    The bottom plot compares the absolute error of both methods. The analytical solution of the Riemann problem is described in full detail in Appendix~\ref{sec:appendix2_shock_tube}. The $L_2$ error for the HLLE solver is $\sim 0.08$, while the Marquina solver performs slightly better with an error of $\sim 0.06$. However, the Marquina solution exhibits spurious oscillations in the orange region, which can be amplified and lead to numerical instability in some cases. For this reason, we adopt the HLLE solver for the subsequent test cases. A more detailed investigation of the numerical instabilities associated with the Marquina solver is left for future work.

\subsection{Black hole accretion}

    Simulating \bh{} in a radiation bath is one of the main motivations for developing the radiation fluid code, and it is essential to test radiation dynamics in the presence of a puncture. In this test, a \bh{} is surrounded by a Gaussian shell of radiation at rest centered at $r_0=8M$, with amplitude $\delta e= 10^{-4} (M^{-2})$ and width $\sigma = 3M$, and the evolution shows how radiation gets accreted to the BH. The computational domain spanning $[0,80]M$ with $6$ levels of mesh refinement with $50$ grid points across each level, with the finest resolution $\Delta=0.05 M$.\\  

    For this spherically symmetric spacetime, we express the metric in isotropic coordinates:
    \begin{equation}
        ds^2 = -\alpha^2(r) dt^2 + \psi^4(r) \left( dr^2 + r^2 d\Omega \right)
    \end{equation}
    To prepare the spacetime profile for this configuration, we use a modified Bowen-York type initial data introduced in~\cite{Clark_2016}. Following the conformal-transverse-traceless (CTT) approach pioneered by Lichnerowicz~\cite{Lichnerowicz1944} and York~\cite{York1971,York:1979} (see Ch.~3 in~\cite{baumgarte_shapiro_2010} for a review), we conformally scale $\gamma_{ij}, \rho_H, S^i$,
    \begin{eqnarray}
        \tilde{\gamma}_{ij} &=&  \gamma_{ij} \psi^{-4}\\
        \tilde{\rho}_H &=& \rho_H \psi^8\\
         \tilde{S}^i &=& S^i \psi^{10}\,.
    \end{eqnarray}
    and decompose the traceless component of the extrinsic curvature $K_{ij}$,
    \begin{equation}
        K_{ij} = \psi^{-2} \tilde{A}_{ij} + \frac{1}{3}\gamma_{ij}K.
    \end{equation}
    By imposing conformal flatness ($\tilde{\gamma}_{ij} = \eta_{ij}$) and maximal slicing ($K=0$), the constraint equations become
    \begin{eqnarray}
        \label{eq:ham2}
        &\tilde{\Delta} \psi+ \frac{1}{8}\psi^{-7}\tilde{A}_{ij}\tilde{A}^{ij} = -2\pi \psi^{-3}\tilde{\rho}_H \\
        \label{eq:mom2}
        &\widetilde{\nabla}_j \tilde A^{ij} = 8 \pi \tilde{S}^i.
    \end{eqnarray}
    
    The momentum constraint Eq.(\ref{eq:mom2}) is now linear and analytical solutions exist for punctures ($\tilde{S}^i=0$) and spherically symmetric objects ($\tilde{S}^i=P^i \sigma (r)$ for a constant vector $P^i$), as derived in~\cite{Bowen1979}. In the \bh{} accretion case, both the BH and the radiation shell are at rest initially, and thus $\tilde{A}_{ij}=0$.

    The Hamiltonian constraint equation Eq.(\ref{eq:ham2}) is solved with a modified \texttt{TwoPunctures}~\cite{Ansorg_2004} code with a source $\rho_H$. The singular term in the conformal factor for a puncture with bare mass $m$ can be decomposed to
    \begin{equation}
        \psi = 1 + \frac{m}{2r}+u,
    \end{equation}
    and $u$ is then solved using spectral methods. In this case, we choose bare mass $m=1.0$, which will give a slightly larger initial \bh{} mass $M_{\rm BH} \gtrsim m$. 
    
    Radiation has no conserved rest mass, so we instead use the Killing energy $E_{\rm rad}$ as the relevant conserved energy in this spacetime,
    \begin{equation}
        E_{\rm rad}
        =
        \int
        T^{\mu\nu} n_\mu t_\nu \sqrt{\gamma}\, d^3x
        =
        \int
        \left(\alpha \rho_H - \beta^i J_i\right)
        \sqrt{\gamma}\, d^3x .
    \end{equation}
    where $\rho_H, J_i$ are the projections of the stress energy tensor in Eq.(\ref{eq:T_proj}). The total energy of the system is thus approximated by
    \begin{equation*}
        M_{ADM} \approx M_{\rm tot}(t) := M_{\rm BH}(t) + E_{\rm rad}(t).
    \end{equation*}

\begin{figure}[!htbp]
        \centering
        \includegraphics[width=0.48\textwidth]{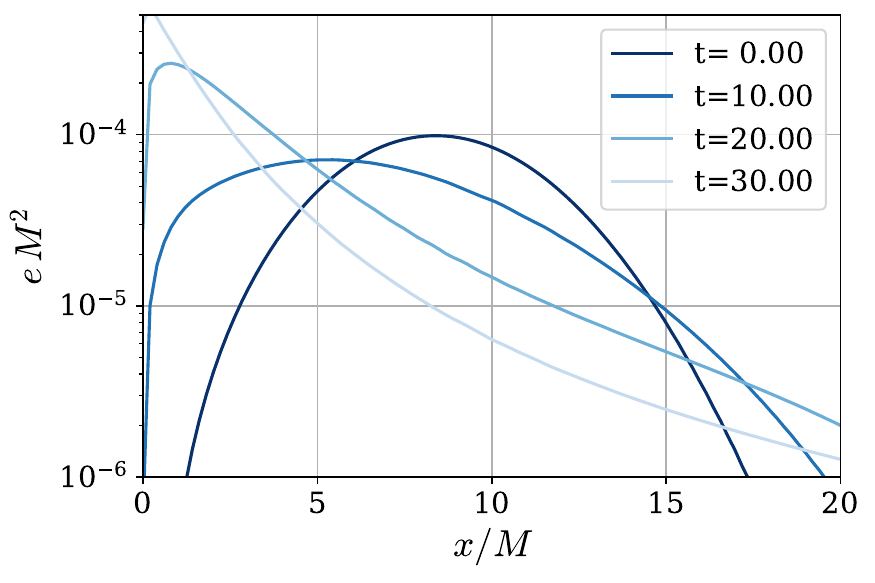}
        \caption{The energy density profile for black hole accretion. The plot shows various snapshots of the energy density profile on the $x$-axis from an initial Gaussian shell at $r=8M$ to clustering near the puncture at the origin.}
        \label{fig:BHRad_1D}
    \end{figure}

\begin{figure}[!htbp]
        \centering
        \includegraphics[width=0.48\textwidth]{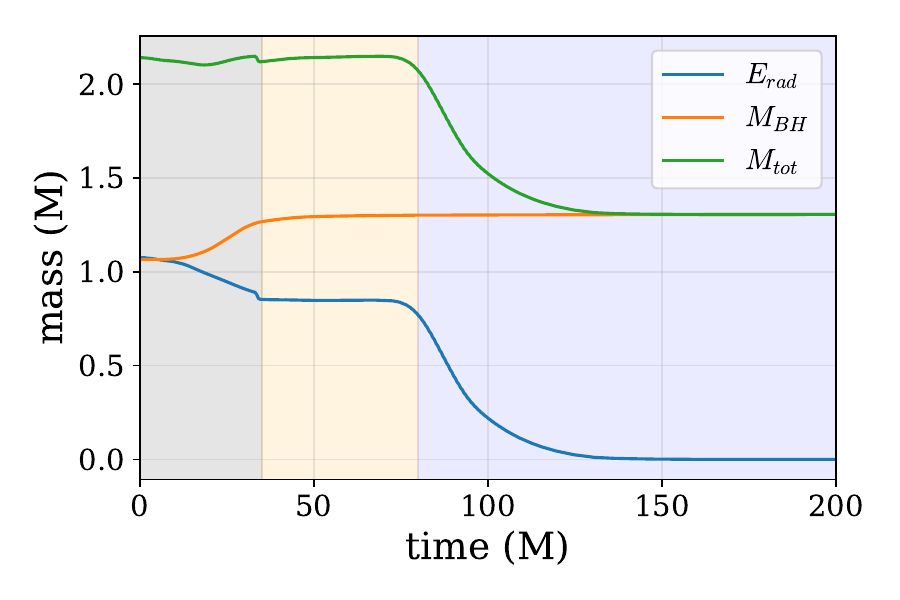}
        \caption{The mass accretion and total radiation energy for black hole accretion. The orange and blue curves are the irreducible mass of the BH $M_{BH}$ and total energy of the radiation $E_{rad}$, respectively. The green curve is the approximate total mass by summing the two $M_{tot} = M_{BH} + E_{rad}$. Three phases of the evolution can be shown: radiation accretion onto the \bh{} $t\in[0,35]M$ (gray background), the propagation of the escaped radiation $t\in[35,80]M$ (orange background), and finally the radiation leaving the computational domain at $t>80 M$ (blue background).}
        \label{fig:BHRad_mass}
    \end{figure}

    The energy density profile for black hole accretion is shown in Fig.~\ref{fig:BHRad_1D}. The initial Gaussian shell at $r=8M$ immediately  accretes to the puncture at the origin within $t=30M$. The mass accretion and total radiation energy for black hole accretion are shown in Fig.~\ref{fig:BHRad_mass}. The mass of the BH, $M_{BH}$ (orange curve) and total energy of the radiation $E_{rad}$ (blue curve) are summed to approximate the total energy of the system (green curve) that remains mostly constant ($M_{tot}\sim2.13M$) before $t=80M$. Three phases of the evolution can be identified: radiation accretion ($\Delta E_{rad}\approx\Delta M_{BH}\sim0.24M$) onto the \bh{} at time $t\in[0,35]M$ (gray background), the propagation of the escaped radiation at time $t\in[35,80]M$ (orange background), and finally, the radiation leaving the computational domain at time $t>80 M$ (blue background) ($\Delta M_{tot}\sim 0.83 M$).

\subsection{Photon star instability}
    \begin{figure*}[!htbp]
            \centering
            \includegraphics[width=0.96\textwidth]{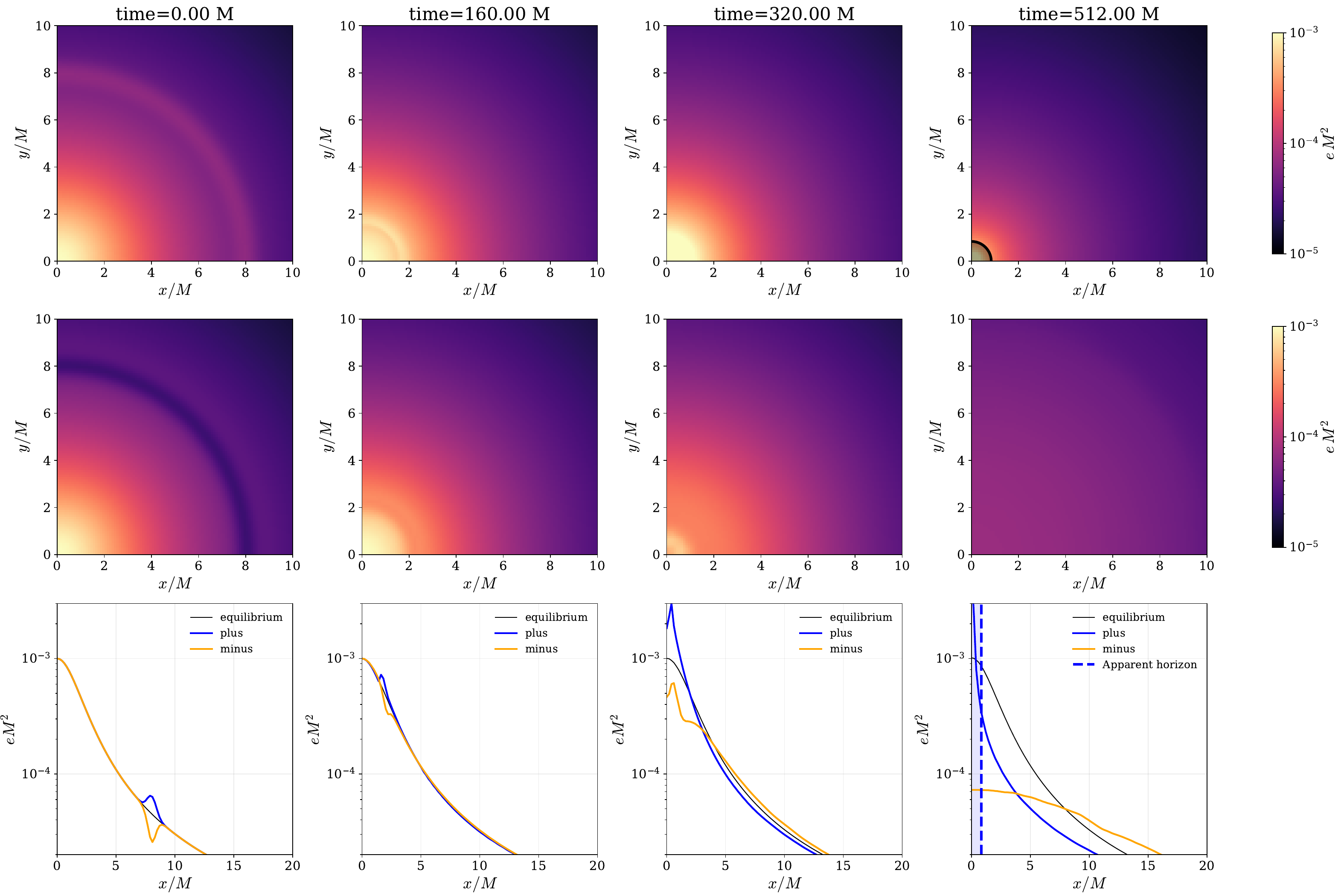}
            \caption{The radiation star solution and its instability under perturbations are shown. The density evolution of the perturbed cases is displayed in the top two rows, while density slices along the x-axis for all three cases are shown in the bottom row. The first row shows the \textit{plus} perturbation propagating toward the center of the star and eventually collapsing to a black hole, with the apparent horizon visible at time 512M, indicated by the black circle in the top panels and the blue vertical line in the bottom panels. The second row shows the \textit{minus} perturbation also reaching the center, where it reflects and subsequently triggers the dispersal of most of the radiation.}
            \label{fig:tov}
    \end{figure*}
    A nontrivial equilibrium spacetime made entirely of radiation, or a ``photon star'', can be constructed by solving the TOV equations. In isotropic coordinates, the TOV equations solve the coupled differential equations:
    \begin{align}
        \frac{1}{r^2}(r^2\psi')' &= -2\pi \psi^5 e \\
        \frac{1}{r^2}(r^2\Theta')' &= 2\pi \Theta \psi^4 (e+6P) \\
        P' = -&(e + P) \left(\frac{\Theta'}{\Theta} - \frac{\psi'}{\psi}\right),
    \end{align}
    where $\Theta(r) \equiv \alpha(r) \psi(r)$.
    
    The pressure is integrated from the central pressure given by central energy density $P_c = e_c/3$. For relativistic stars with $\Gamma=4/3$, the solution gives $e \rightarrow{} r^{-2}$ as $r \rightarrow{}\infty$ instead of reaching a surface when $e=0$, resulting in an infinite mass with non-vanishing density everywhere~\cite{Chavanis:2007kn}.  For baryon fluids, typical treatments for integrating $\psi,\Theta$ starts both at $1.0$ and rescaled all TOV variables given the conformal factor at the surface of the star $r=R_*$, such that the exterior gives $\psi=1+M/2r$ and $\Theta=1-M/2r$ given by the total mass of the star $M=\int_0^{R_*}4\pi r^2 \psi^5e\ dr$. However, since there is no surface for photon stars, the rescaling is done by setting $\psi\sim\Theta\sim1$ at $r\rightarrow\infty$ for a chosen radius bigger than our computational domain.\\
    
    The stability analysis of photon stars typically encloses a domain with a finite radius $R$. These equilibrium configurations are dynamically unstable, such that perturbations are expected to drive either collapse or expansion~\cite{Evans:1994pj}. Therefore, we perform three runs: \textit{equilibrium, plus, minus} to show the cases of unperturbed, added and subtracted perturbations for the initial energy density profile.\\
    
    In this test case, we choose central energy density $e_c=0.001 (M^{-2})$ and a computational domain spanning $[0,160]M$ with $5$ levels of mesh refinement with $50$ grid points across each level, with the finest resolution $\Delta=0.2M$.  The energy density perturbation $\delta e$ are introduced with a Gaussian bump with $\delta e_0 = 0.02e_c$, $r_0=8M$, and $\sigma\sim0.353 M$. The perturbation is then either added and subtracted from the initial energy density profile for the \textit{plus} and \textit{minus} cases to assess the dynamical instability under perturbation. To resolve the apparent horizon of the BH formed in the \textit{plus} case, two additional levels are used, lowering the finest resolution to $\Delta_+=0.05M$.\\
    
    The results of the three cases are shown in Fig.~\ref{fig:tov}. As shown in the first row and the blue curve in the third row, the added perturbation travels toward the center of the star initially, and entire configuration collapses to a \bh{} at time $t=512M$. The subtracted perturbation, shown in the second row and the orange curve in the third row, travels toward the center and an initial implosion occurs, but the energy peak is reflected and immediately triggers a blast wave sending most of the energy outward. The results agree with the predictions from stability analysis such that perturbations either drive collapse or dispersion from the unstable equilibrium.\\

\section{Conclusion}
    \label{sec:Conclusions}
    We have presented a formulation for evolving radiation fluids within the general relativistic hydrodynamics framework of the Einstein Toolkit by modeling radiation as a self-gravitating perfect fluid with vanishing rest mass. By modifying the source terms, recovery scheme, and Riemann solvers in the existing \whisky code, we have demonstrated that standard high-resolution shock-capturing methods can be extended to handle radiation fluids, and validate the implementation through shock tube tests. As an application, we model radiation accretion onto a black hole to demonstrate the robustness of the method in the presence of punctures. In addition, we examine the evolution of photon star initial data obtained from the TOV solution and confirm its dynamical instability under perturbations, leading to either collapse or dispersal. These results establish an effective framework for studying self-gravitating radiation in strong-gravity spacetimes and provide a foundation for future extensions to other radiation-dominated spacetimes including singular TOV solutions.\\

\section*{Data availability}
    Data and code supporting the findings of this study are available from the corresponding author upon reasonable request.\\

\section*{Acknowledgments}
    \label{sec:Ackowledgement}
    This work is supported by NSF grants PHY-2411068 and PHY-2207780. We would like to thank Deirdre Shoemaker, Miguel Gracia-Linares and Jake Doherty for their helpful comments. The authors used ChatGPT-6 and Claude (Opus 5.5) to review the manuscript for language and clarity and GitHub Copilot to assist with code generation. The usage and outputs of AI tools are carefully reviewed by the authors.\\

\bibliographystyle{apsrev4-2}
\bibliography{References}

\appendix
\onecolumngrid
\section{Marquina Riemann Solver for radiation}
\label{sec:appendix1_riemann}
    The Marquina Riemann solver for standard SRHD~\cite{Donat_1998,Aloy_1999} and GRHD~\cite{ibanez_1999} uses the characteristic decomposition of the flux Jacobian. In this section, we show the modified eigenvalues and eigenvectors for the radiation fluid $4\times 4$ Jacobian matrix $\partial \mathbf{F}^x / \partial \mathbf{U}$ in the $x$-direction for $\mathbf{U}=(\tau,S_x,S_y,S_z)$.\\

    For the general-relativistic case, we follow similar procedure as~\cite{ibanez_1999} and find the same eigenvalues. 
    \begin{equation}
        \lambda_\pm = \alpha \frac{
    (1 - c_s^2)\, v^x\ \pm \sqrt{c_s^2\,\bigl(1 - v^2\bigr)\,
    \bigl(\gamma^{xx} (1-v^2 c_s^2) - v^xv^x(1 - c_s^2)\bigr)}}{1 - c_s^2 v^2} - \beta^x
    \end{equation}
    
    \begin{equation}
         \lambda_0 = \alpha v^x - \beta^x \quad (\text{double}).
    \end{equation}
    where $v^2 = v_x v^x + v_y v^y + v_z v^z$ and the upper spatial metric $\gamma^{ij} = (\gamma_{ij})^{-1}$.
    
    To derive the eigenvectors, we first define
    \begin{equation*}
        \Lambda_\pm = \frac{\lambda_\pm}{\alpha} + \frac{\beta^x}{\alpha} = \frac{
        (1 - c_s^2)\, v^x\ \pm \sqrt{c_s^2\,\bigl(1 - v^2\bigr)\,
        \bigl(\gamma^{xx} (1-v^2 c_s^2) - v^xv^x(1 - c_s^2)\bigr)}}{1 - c_s^2 v^2}
    \end{equation*}

    The right eigenvectors $r_{(p)}$ are:
    \begin{equation}
        r_{+} = 
        \begin{pmatrix}
            \gamma^{xx} A_{+} \\
            A_{+}\Lambda_{+} - (v_y \gamma^{xy} + v_z \gamma^{xz}) \\
            \gamma^{xx} v_y \\
            \gamma^{xx} v_z
        \end{pmatrix},
        \qquad
        r_{-} =
        \begin{pmatrix}
            \gamma^{xx} A_{-} \\
            A_{-}\Lambda_{-} - (v_y \gamma^{xy} + v_z \gamma^{xz}) \\
            \gamma^{xx} v_y \\
            \gamma^{xx} v_z
        \end{pmatrix}.
    \end{equation}
    \begin{equation}
        r_{0,1} =
        \begin{pmatrix}
            \dfrac{2(\gamma^{xx} v^y - \gamma^{xy} v^x)}
                  {\gamma^{xx}(1+v^2) - 2 v^x v^x} \\[8pt]
            \dfrac{2 v^x v^y - \gamma^{xy}(1+v^2)}
                  {\gamma^{xx}(1+v^2) - 2 v^x v^x} \\[8pt]
            1 \\[4pt]
            0
        \end{pmatrix},
        \qquad
        r_{0,2} =
        \begin{pmatrix}
            \dfrac{2(\gamma^{xx} v^z - \gamma^{xz} v^x)}
                  {\gamma^{xx}(1+v^2) - 2 v^x v^x} \\[8pt]
            \dfrac{2 v^x v^z - \gamma^{xz}(1+v^2)}
                  {\gamma^{xx}(1+v^2) - 2 v^x v^x} \\[8pt]
            0 \\[4pt]
            1
        \end{pmatrix}.
    \end{equation}
    
    The left eigenvectors $\mathbf{\ell}_{(p)}$ are:
    \begin{equation}
        \ell_{0,1} =
        \begin{pmatrix}
            - v_y k_x \\[6pt]
            v^x v_y k_x \\[6pt]
            1 + \dfrac{2 v^y v_y}{1 - v^2}
            - \dfrac{v^x \gamma^{xy} v_y}{\gamma^{xx} - v^x v^x} \\[8pt]
            \dfrac{2 v^z v_y}{1 - v^2}
            - \dfrac{v^x \gamma^{xz} v_y}{\gamma^{xx} - v^x v^x}
        \end{pmatrix},
        \qquad
        \ell_{0,2} =
        \begin{pmatrix}
            - v_z k_x \\[6pt]
            v^x v_z k_x \\[6pt]
            \dfrac{2 v^y v_z}{1 - v^2}
            - \dfrac{v^x \gamma^{xy} v_z}{\gamma^{xx} - v^x v^x} \\[8pt]
            1 + \dfrac{2 v^z v_z}{1 - v^2}
            - \dfrac{v^x \gamma^{xz} v_z}{\gamma^{xx} - v^x v^x}
        \end{pmatrix}.
    \end{equation}

    \begin{equation}
        \ell_{+} =
        \dfrac{1}{\gamma^{xx}}
        \begin{pmatrix}
            \dfrac{v^x}{2 A_{+} (v^x - \Lambda_{+})} + \dfrac{k_x}{2} \\[4pt]
            - \left( \dfrac{\gamma^{xx}}{2 A_{+} (v^x - \Lambda_{+})} + \dfrac{k_x v^x}{2} \right) \\[4pt]
            - \dfrac{v^y}{1 - v^2} + \gamma^{xy}\dfrac{(C_{1x} \Lambda_{+} + v^x C_{2x}) }{2 (\gamma^{xx} - v^x v^x)} \\[4pt]
            - \dfrac{v^z}{1 - v^2} + \gamma^{xz}\dfrac{(C_{1x} \Lambda_{+} + v^x C_{2x})}{2 (\gamma^{xx} - v^x v^x)}
        \end{pmatrix},
        \qquad
        \ell_{-} =
        \dfrac{1}{\gamma^{xx}}
        \begin{pmatrix}
            \dfrac{v^x}{2 A_{-} (v^x - \Lambda_{-})} + \dfrac{k_x}{2} \\[4pt]
            - \left( \dfrac{\gamma^{xx}}{2 A_{-} (v^x - \Lambda_{-})} + \dfrac{k_x v^x}{2} \right) \\[4pt]
            - \dfrac{v^y}{1 - v^2} + \dfrac{(C_{1x} \Lambda_{-} + v^x C_{2x}) \gamma^{xy}}{2 (\gamma^{xx} - v^x v^x)} \\[4pt]
            - \dfrac{v^z}{1 - v^2} + \dfrac{(C_{1x} \Lambda_{-} + v^x C_{2x}) \gamma^{xz}}{2 (\gamma^{xx} - v^x v^x)}
        \end{pmatrix}.
    \end{equation}

    where
    \begin{equation*}
        A_{+} = \frac{\gamma^{xx} - v^x v^x}{\,\gamma^{xx} - v^x \, \Lambda_{+}\,}, 
        \qquad
        A_{-} = \frac{\gamma^{xx} - v^x v^x}{\,\gamma^{xx} - v^x \, \Lambda_{-}\,},
    \end{equation*}
    \begin{equation*}
        C_{1x} = \frac{1 - c_s^2 v^2}{c_s^2 (1 - v^2)},
        \qquad
        C_{2x} = 1 - \frac{1 - c_s^2}{c_s^2 (1 - v^2)},
        \qquad
        k_x = \frac{\gamma^{xx}(1+v^2) - 2 v^x v^x}{(\gamma^{xx}- v^x v^x)(1 - v^2)}.
    \end{equation*}\\

    The characteristic decomposition for the radiation fluid in this section provides the structure needed for the Riemann problem in our Cartesian grid. However, extending to curvilinear coordinates would not be trivial. The quasi-invertible transformation method developed for relativistic hydrodynamics~\cite{Teukolsky:2025erz, Teukolsky:2025yle} provides a more systematic approach and could be explored in future work.

\section{Shock tube test analytical solution}
\label{sec:appendix2_shock_tube}
    Here we derive the analytical solution for the 1D shock tube test for radiation fluid as benchmark for the Riemann solver test case.
    \[
        e(x,0)=
        \begin{cases}
        e_L, & x<0,\\
        e_R, & x>0,
        \end{cases}
        \qquad
        v_L=v_R=0,
        \qquad
        P=\frac{e}{3}, \qquad c_s = \sqrt{\frac{1}{3}}.
    \]
    
    Assume \(e_L>e_R\). Then the solution is described by Sec.~5 in~\cite{Rezzolla:2001kn} to have one shock and one rarefaction wave:
    \[
    L \ | \ \text{left rarefaction} \ | \ * \ | \  \text{right shock} \  | \ R.
    \]
    
    Across the left rarefaction, the Riemann invariant gives
    \begin{equation}
        \label{eq:shock_sol_v1}
        v_*=
        \tanh\left[
        \frac{\sqrt{3}}{4}
        \ln\left(\frac{e_L}{e_*}\right)
        \right].
    \end{equation}
    Across the right shock,
    \begin{equation}
        \label{eq:shock_sol_v2}
        v_*
        =
        \sqrt{\frac{c_s^2\,(e_* - e_R)^2}
        {c_s^2 e_*^2 + (1 + c_s^4)e_* e_R + c_s^2 e_R^2}}.
    \end{equation}
    Combining the two equations will give $e_*, v_*$.\\
    
    The speed of the rarefaction head and tail are
    \begin{equation}
        \label{eq:shock_sol_lambda}
        \lambda_{\rm head}
        =
        \frac{v_L-c_s}{1-v_Lc_s}
        ,\qquad 
        \lambda_{\rm tail}
        =
        \frac{v_*-c_s}{1-v_*c_s}
    \end{equation}
    
    The shock speed follows from
    \begin{equation}
        \label{eq:shock_sol_s}
        s = \frac{T_*^{0x}-T_R^{0x}}{T_*^{00}-T_R^{00}},
    \end{equation}
    where
    \[
    T^{00}=\frac{4}{3}eW^2-\frac{1}{3}e,
    \qquad
    T^{0x}=\frac{4}{3}eW^2v,
    \qquad
    W=\frac{1}{\sqrt{1-v^2}}.
    \]

    For \(e_L=10\) and \(e_R=1\), Eq.~\eqref{eq:shock_sol_v1} and \eqref{eq:shock_sol_v2} give the energy density and velocity in the star region
    \begin{equation}
        \label{eq:shock_sol_ev_num}
        e_* \simeq 3.1398,
        \qquad
        v_* \simeq 0.4634.
    \end{equation}
    Eq.~\eqref{eq:shock_sol_lambda} and \eqref{eq:shock_sol_s} give,
    \begin{equation}
        \label{eq:shock_sol_lam_s_num}
        \lambda_{\rm head}\simeq -0.5774,
        \qquad
        \lambda_{\rm tail}\simeq -0.1556,
        \qquad
        s\simeq 0.7521.
    \end{equation}
    Finally, in the rarefaction region, given \(v_L=0\), the energy density and velocity can be given as a function of $\xi=x/t$:
    \begin{equation}
        v(\xi)
        =
        \frac{\xi + c_s}
        {1 + \xi c_s}, \quad 
        e(v)
        =
        e_L
        \exp\left(
        -\frac{4}{\sqrt{3}}\tanh^{-1}(v)
        \right).
    \end{equation}
    
\end{document}